\documentclass{aa}
\usepackage{natbib}
\usepackage{epsfig}
\usepackage{txfonts}

\def\Msol{M$_\odot$}             %Msun
\def\Rsol{R$_\odot$}             %Msun
\def\Ofive{OGLE-TR-56$b$\ }
\def\Oten{OGLE-TR-10$b$\ }

\begin{document}
%%%%%%%%%%%%%%%%%%%%%%%%%%%%%%%%%%%%%%%%%%%%%%%%%%%%
\title{The `666' collaboration on OGLE transits: I.  Accurate radius of the planets OGLE-TR-10$b$ and OGLE-TR-56$b$ with VLT deconvolution photometry\thanks{Based on data collected with the FORS1 imager at the VLT-Kueyen telescope (Paranal Observatory, ESO, Chile)  in the programme 177.C-0666E.}}

\author{F. Pont$^1$, C. Moutou$^2$,  M. Gillon$^{1, 3}$,  A. Udalski$^{10}$,  F. Bouchy$^{5, 6}$, J. M. Fernandes$^{14}$, W.  Gieren$^{7}$,  M. Mayor$^1$,  T. Mazeh$^8$, D. Minniti$^4$, C. Melo$^9$, D. Naef$^9$, G. Pietrzynski$^{7, 10}$, D. Queloz$^1$, M. T. Ruiz$^{11}$, N.C. Santos$^{1, 12, 13}$,  S. Udry$^1$}

\offprints{frederic.pont@obs.unige.ch}
\institute{$^1$  Observatoire de Gen\`eve, 51 Chemin des Maillettes,1290 Sauverny,
    Switzerland\\
$^2$ Laboratoire d'Astrophysique de Marseille, Traverse du Siphon, BP8, Les Trois Lucs, 13376 Marseille
  cedex 12, France\\    
$^3$ Institut d'Astrophysique et de G\'eophysique,  Universit\'e
  de Li\`ege,  All\'ee du 6 Ao\^ut 17,  4000 Li\`ege, Belgium \\
$^4$ Departmento de Astronom\'ia y Astrof\'isica, Pontificia Universidad Cat\'olica de Chile, Casilla 306, Santiago 22, Chile\\
$^5$ Observatoire de Haute-Provence, 04870 St-Michel l'Observatoire, France\\
$^6$ Institut d'Astrophysique de Paris, 98bis Bd Arago, 75014 Paris, France\\
$^7$ Departamento de Fisica, Astronomy Group, Universidad de Concepci\'on, Casilla 160-C, Concepci\'on, Chile\\
$^8$ School of Physics and Astronomy, Raymond and Beverly Sackler Faculty of Exact Sciences, Tel Aviv University, Tel Aviv, Israel\\
$^9$ European Southern Observatory, Casilla 19001, Santiago 19, Chile\\
$^{10}$ Warsaw University Observatory, Al. Ujazdowskie 4, 00-478, Warsaw, Poland\\
$^{11}$ Department of Astronomy, Universidad de Chile, Santiago, Chile\\
$^{12}$ Centro de Astronomia e Astrof{\'\i}sica da Universidade de Lisboa,
  Observat\'orio Astron\'omico de Lisboa, Tapada da Ajuda, 1349-018
  Lisboa, Portugal\\
$^{13}$   Centro de Geofisica de \'Evora, Rua Rom\~ao Ramalho 59,
  7002-554 \'Evora, Portugal\\
  $^{14}$ Grupo de Astrof\'isica UC, Observat\'orio Astron\'omico da Universidade de 
Coimbra, Santa Clara, Coimbra, Portugal\\}
\date{Received date / accepted date}

   \authorrunning{F. Pont et al.}
   \titlerunning{Radius of the transiting planets OGLE-TR-10$b$ and OGLE-TR-56$b$}
%%%%%%%%%%%%%%%%%%%%%%%%%%%%%%%%%%%%%%%%%%%%%%%%%%%%
\abstract{Transiting planets are essential to study the structure and evolution of extra-solar planets. For that purpose, it is important to measure precisely the radius of these planets. Here we report new high-accuracy photometry of the transits of OGLE-TR-10 and OGLE-TR-56 with VLT/FORS1. One transit of each object was covered in Bessel $V$ and $R$ filters, and treated with the deconvolution-based photometry algorithm DECPHOT, to ensure accurate millimagnitude light curves.  Together with earlier spectroscopic measurements, the data imply a radius of 1.22 $^{+0.12}_{-0.07}$ R$_J$ for OGLE-TR-10$b$ and 1.30 $\pm 0.05$ R$_J$  for OGLE-TR-56$b$. A re-analysis of the original OGLE photometry resolves an earlier discrepancy about the radius of OGLE-TR-10. The transit of OGLE-TR-56 is almost grazing, so that small systematics in the photometry can cause large changes in the derived radius. Our study confirms both planets as inflated hot Jupiters, with large radii comparable to that of HD 209458$b$ and at least two other recently discovered transiting gas giants.

}
\maketitle
%%%%%%%%%%%%%%%%%%%%%%%%%%%%%%%%%%%%%%%%%%%%%%%%%%%%

\section{Introduction}

\begin{figure*}[thbp]
\centering                     
\includegraphics[width=12.0cm]{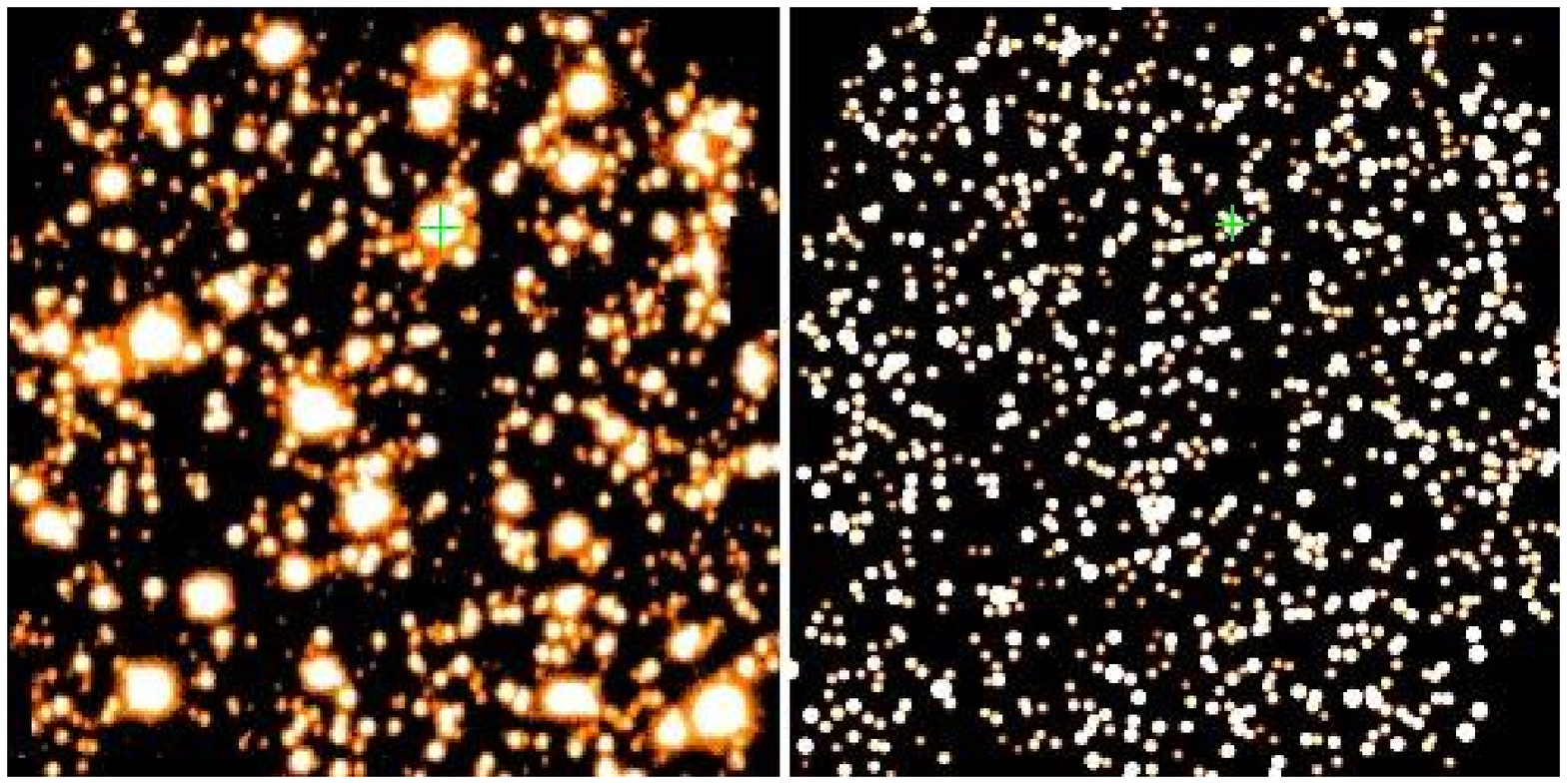}\includegraphics[width=6.0cm]{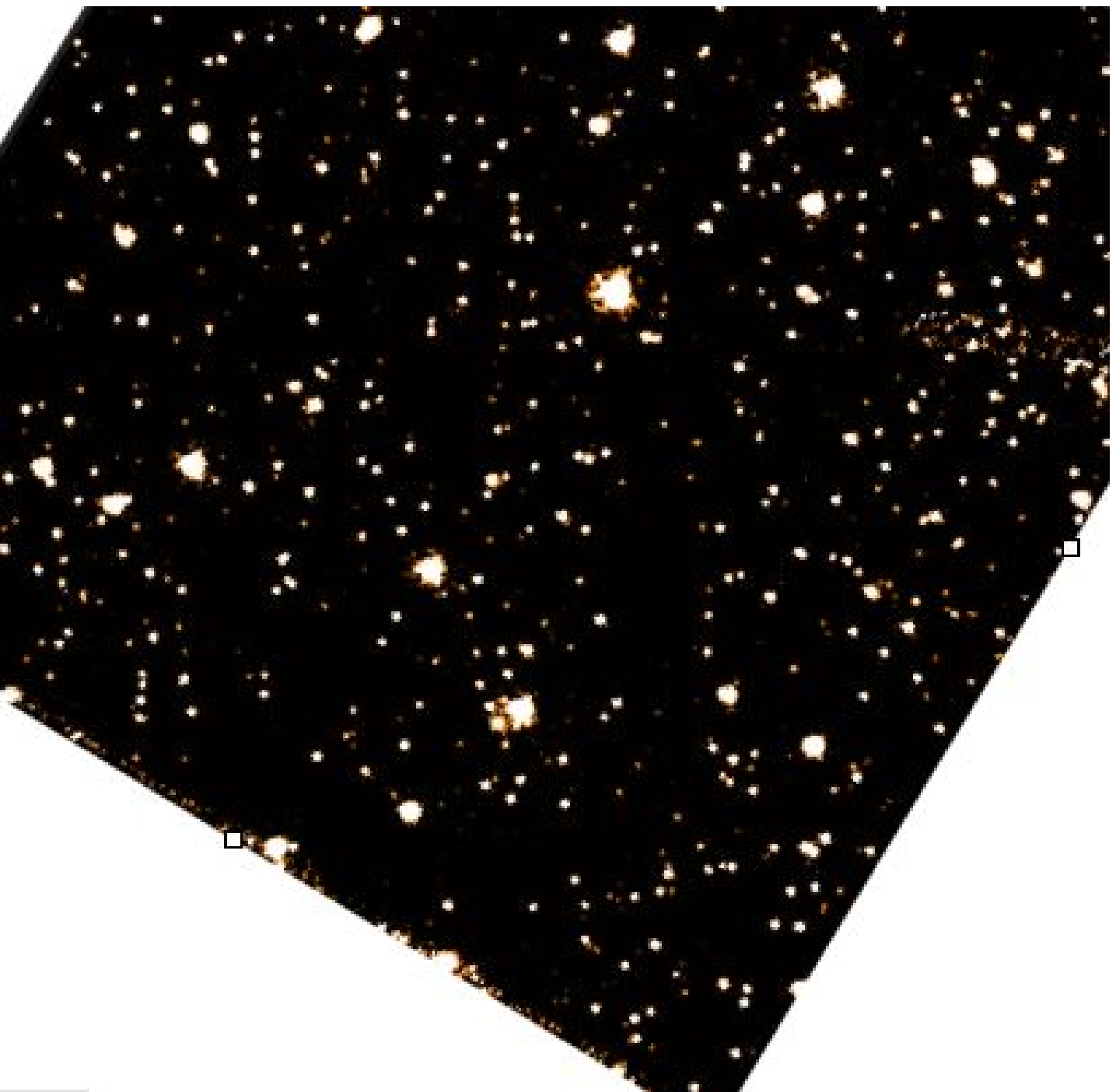}
\caption{\emph{Left}: OGLE-TR-56 (marked with a cross) in a 256 pixels $\times$ 256 pixels sub-image (0.51 $\arcmin$ $\times$ 0.51 $\arcmin$) from the best seeing VLT/FORS1 image of the run ($top$ = North, $left$ = East, $FWHM$ = 0.4\arcsec, some areas at the edges were masked for the reduction). \emph{Middle}: the same image after deconvolution ($FWHM$ = 0.2\arcsec). \emph{Right}: an HST image of the same field. These images illustrate the high level of crowding in the fields of view of the OGLE-III survey towards the Galactic Bulge, and the capacity of deconvolution to recover the information.}
\label{fig:a}
\end{figure*}

Fourteen transiting extrasolar planets have been detected so far\footnote{For an up-to-date list see e.g.\\ obswww.unige.ch/~pont/TRANSITS.html.}, and they play a central role in the study of the structure and evolution of hot Jupiters (Charbonneau 2005). Among them, \Ofive was the first planet detected by a photometric transit survey (Udalski et al. 2002b) then confirmed by radial velocity measurements (Konacki et al. 2003), and also the prototype of a new group of planets revealed by transit surveys, the ``Very Hot Jupiters''. This planet was unique because it orbited closer to its star than any other known planet at that time. Indeed, no planet with a lower orbital period has been confirmed so far\footnote{There are some possible candidates, e.g. Fern\'andez et al. (2006)}. New spectroscopic data have since been added by Torres et al. (2004), Bouchy et al. (2005) and Santos et al. (2006) to refine the velocity orbit and spectroscopic parameters, but OGLE-TR-56 has not been re-observed with higher-accuracy photometry. Photometric observations were obtained with the Hubble Space Telescope using the ACS camera (proposal GO/9805), without conclusive results.

\Oten was also detected by the OGLE-III transit survey (Udalski et al. 2002a) and confirmed later by radial velocity measurements (Bouchy et al. 2005; Konacki et al. 2005). By combining the spectroscopic information on the host star with the depth and shape of the photometric transit, the size of the planet orbiting OGLE-TR-10 could be estimated.

However, the issue of the radius of \Oten  has proved vexing. The earliest calculated radius (1.54 $\pm 0.12 R_J$, Bouchy et al. 2005) suggested a highly inflated planet, the largest detected so far, with a very low density and important implications for the theory of planetary structure and formation. Subsequently however, Konacki et al. (2005) found a lower radius of 1.24 $\pm 0.09$ R$_J$, due to the fact that they obtained a lower temperature for the host star. This radius value still implied a bloated hot Jupiter, but not pathologically large. Then Holman et al. (2006, hereafter H06) announced more precise photometric measurements that indicated a much shallower transit than found from the OGLE photometry, correspondingly bringing down the computed radius to 1.06 R$_J$. But Santos et al. (2006), with new high signal-to-noise high resolution spectra for OGLE-TR-10, obtained a higher temperature than indicated by Konacki et al. (2005) and H06, and a significantly super-solar metallicity of the star. These parameters implied a radius of 1.14 $R_J$ if the H06 photometry is used, and 1.43 R$_J$ with the original OGLE photometry. Table~\ref{tab:d} summarizes the evolution of radius estimates for \Oten in the past two years.

Meanwhile Gillon et al. (2007), in a recent re-analysis of a transit of OGLE-TR-132$b$ observed with the VLT earlier (Moutou et al. 2004) outlined a normalisation problem with some implementations of difference image photometry  (the ISIS image subtraction package of Alard 1999 combined with aperture photometry), liable to lead to an underestimation of the transit depth and an overestimation of the accuracy achieved. That analysis showed that the problem was clearly present in the VLT photometry presented by Moutou et al. (2004), and therefore could also be present in the H06 results, obtained with the same reduction method (ISIS+aperture). This suggested the  possibility that part of the mismatch between the depth of the transit given by the OGLE photometry and by H06 be due to reduction-dependent photometric systematics. 

This conundrum provided the motivation for our new photometric data and re-analysis of OGLE-TR-10. The objective was to settle the issue of the transit depth by using another reduction method than differential image analysis on independent data, as well as to re-assess the spectroscopic parameters. For this purpose, we have gathered a new VLT/FORS photometric transit lightcurve and reduced it with a deconvolution method that is especially robust to normalisation biases, in the context of the ESO Large Programme 666 on OGLE transits\footnote{The `666' collaboration (programme 177.C-0666) is devoted to the spectroscopic and photometric follow-up of the transit candidates and transiting planets provided by the OGLE transit survey, using the FLAMES multifiber spectroscope and the two FORS cameras at the VLT.}. 

We observed a transit of OGLE-TR-56 and a partial transit of OGLE-TR-10 in two colours with FORS1 on the VLT and reduced the data with the deconvolution-based photometry software DECPHOT, a reduction method able to perform high accuracy photometry in very crowded fields (see Gillon et al. 2006; Magain et al. 2006). We have also re-reduced part of the H06 data. Using these new photometric data together with previously published data, we derive new parameters for the OGLE-TR-56 and OGLE-TR-10 star-planet systems, with particular emphasis on the planetary radii. The methods used to derive these stellar and planetary parameters are the same as in the following studies of OGLE transiting planets: Bouchy et al. (2004; 2005), Pont et al. (2004; 2005), Moutou et al. (2004), Santos et al. (2006) and Gillon et al. (2006). The reader is referred to those papers for details

\section{Observation and reduction}

\subsection{The data}

\begin{itemize}
\item\emph{OGLE-TR-10$b$:} VLT observations were obtained on June 4th, 2006 on the FORS1 instrument (programme 177.C-0666E). Scheduling constraints prevented us from observing the full transit, and only the second half is present in our data, from Julian date 2453890.656 to 2453890.819. 152 exposures were acquired in a 3.4\arcmin $\times$ 3.4\arcmin\ field of view, in a total execution sequence of 3.9 hours. The  pixel size is 0.1\arcsec. We made alternative sequences of 7-8 images in the Bessel $R$ and $V$ filters, to get a simultaneous two-colour lightcurve. This corresponds to a filter exchange every 10 minutes. During the full sequence, the measured seeing varies between 1.0$\arcsec$ and 1.6$\arcsec$. The airmass of the field decreases from 1.12 to 1.00 then grows to 1.11. The transparency was high and stable.

\item\emph{OGLE-TR-56$b$:} VLT observations were obtained on July 20th, 2006 on the FORS1 camera (programme 177.C-0666E). 136 exposures were acquired with the same observational strategy as for OGLE-TR-10$b$,  in a total execution sequence of 3 hours, from Julian date 2453936.044 to 2453936.976. The measured seeing varies between 0.4\arcsec and 0.7\arcsec. The air mass of the field decreases from 1.08 to 1.00 then grows to 1.06 during the sequence. The transparency was again high and stable.
\end{itemize}
The frames were debiassed and flatfielded with the standard ESO pipeline. 

\subsection{Reduction}

To reduce these new VLT data, we used the deconvolution-based  photometric reduction method DECPHOT, described in  Gillon et al. (2006) and Magain et al. (2006). As explained in these papers, DECPHOT relies on the partial deconvolution of a set of images to the same higher resolution, and allows the detection of faint blended sources undetected in the original images and a very accurate determination of the point-spread function and the photometry without relying on the presence of any isolated star in the field. As OGLE-TR-10 and OGLE-TR-56 lie in  highly crowded fields of the Galactic Bulge (see Fig.~\ref{fig:a}), the minimisation of the systematic effects due to seeing variations along the run is important to reach a high photometric accuracy, and DECPHOT is well suited to perform this task.

\begin{figure}[t!]
\centering                     
\includegraphics[width=9.0cm]{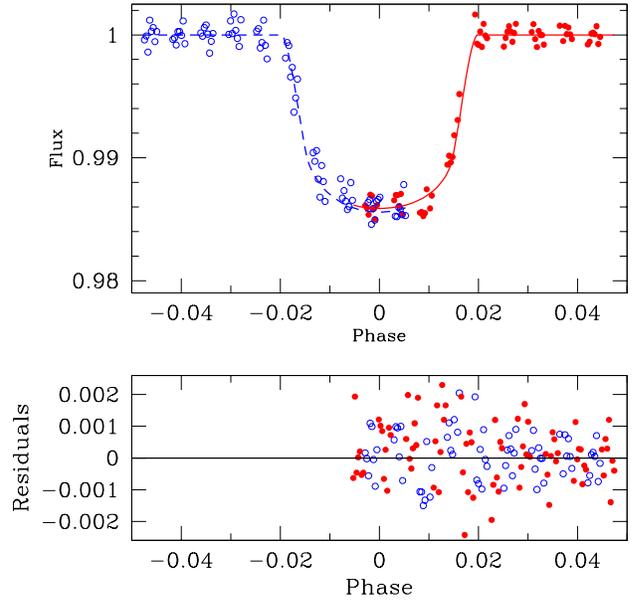}
\caption{VLT photometric data for OGLE-TR-10, in R (filled symbols) and V (open symbols). The model transit curve is shown as solid line for R and dashed line for V. The V data and model curve have been flipped across the transit central epoch for display. The lower panel shows the residuals around the model curve.}
\label{fig:vlt10}
\end{figure}

\begin{figure}[t!]
\centering                     
\includegraphics[width=9.0cm]{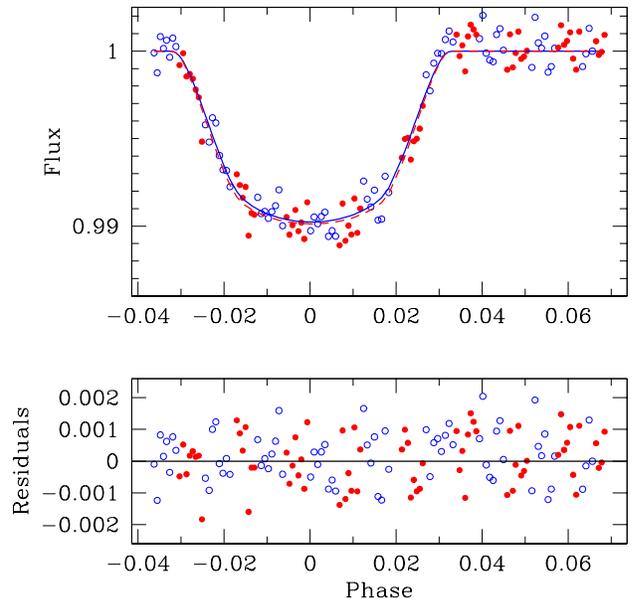}
\caption{VLT photometric data for OGLE-TR-56, symbols as in Fig.~\ref{fig:vlt10}.}
\label{fig:vlt56}
\end{figure}

Figure~\ref{fig:vlt10} presents the light curves obtained  in $V$ and $R$ filter for OGLE-TR-10. Despite the crowding, the standard deviation of the OGLE-TR-10 $V$-band light curve  after the transit is 0.9 mmag (mean photon noise $\sim$ 0.7 mmag), while the deviation of the $R$-band light curve is 0.7 mmag (mean photon noise $\sim$ 0.5 mmag). This demonstrates the high photometric accuracy which can be reached with VLT deconvolution photometry.  Figure~\ref{fig:vlt56} presents the light curves obtained for OGLE-TR-56. Here again, the accuracy is very good: the deviation of the $V $ and $R$ residuals is 0.9 mmag (mean photon noise $\sim$ 0.9 mmag).

\section{Results} 
\subsection{Transit lightcurve fit}
\label{trafit}

The transit shape fitting  was performed with  transit curves computed with the
procedure of Mandel \& Agol (2002), using quadratic limb-darkening
coefficients from Claret (2000).

For OGLE-TR-56, our transit fitting gives $b$ =  0.80 $\pm$  0.02 for the impact parameter, $r/R =0.101 \pm 0.002$ for the radius ratio and $a/R=3.84 \pm 0.16$ for the system scale. This corresponds to a high-latitude  transit, implying a larger primary and larger planet than previous determinations using the original OGLE photometry.

\begin{figure}[t]
\centering                     
\includegraphics[width=9.0cm]{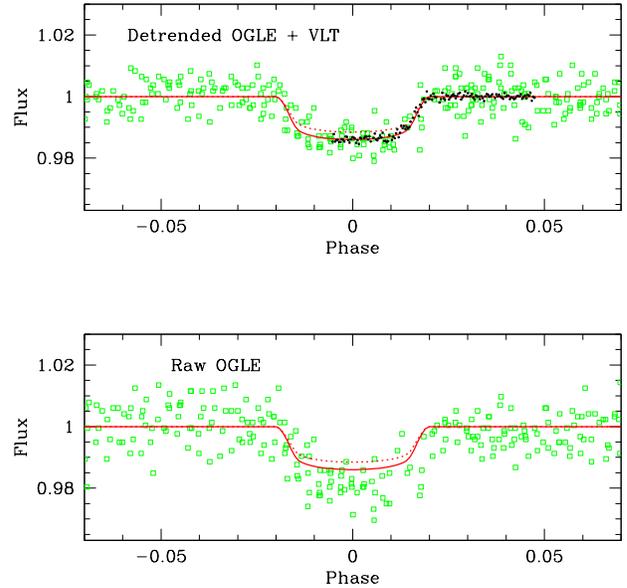}
\caption{\emph{Top}: for OGLE-TR-10, {\it detrended} OGLE photometry (open squares) and VLT/FORS photometry (dots), folded at the best-fit period. The solid line is the model transit from this article, the dashed line from H06. \emph{Bottom}: same with the original OGLE data, prior to detrending.}
\label{ogle10}
\end{figure}

Because only one half of the transit was measured in the case of OGLE-TR-10, we combined our data with the light curve of 1st August 2005 on the Magellan 6.5 telescope from H06, the only night with an entire coverage of the transit with accuracy comparable to our VLT data. H06 have used the differential image analysis code of Alard (1999) in their photometric reduction. They obtain a significantly shallower transit than indicated by our data, by about 20\% (see Fig.~\ref{ogle10}). As mentioned in the Introduction, photometric systematics can affect the scale of a transit lightcurve. Therefore we re-reduced the Magellan images -- kindly provided to us by M. Holman -- with aperture photometry, a method that gives slightly less accurate individual flux measurements, but is more robust against global scale shifts. Our results for the depth of the transit are in perfect agreement with our VLT curve. On closer inspection, it turns out that the cause of the initial mismatch is mainly due to the  measurement of the mean flux outside the transit. The H06 data from 1st August 2005 contains only a short stretch outside the transit, visibly affected by systematics in the differential image reduction. The other nights of data in H06 are not precise enough for a secure definition of the transit depth at the millimagnitude level.

We then fit a transit lightcurve model to our VLT data for OGLE-TR-10 combined with the Magellan data of 1st August 2005. The two datasets yield marginally compatible values for the impact parameter, the first favouring $b \sim 0.55 \pm 0.2$ and the second $0 < b < 0.34$. As shown for instance in Pont \& Moutou (2006) and Bakos et al. (2006), impact parameter and primary radius are almost degenerate in transit lightcurves, and even tiny photometric systematics can amplify the uncertainties on $b$. We use the method of Pont, Zucker \& Queloz (2006) to account for the systematics in the photometry. Instead of using the $\Delta\chi^2 = \chi_{min}^2+1$ contour to define the 1-$\sigma$ error, we use $\Delta\chi^2=\chi^2_{min}+1+\Delta\chi^2_{syst}$, with $\Delta\chi^2_{syst}=n\sigma_r^2/\sigma_w^2$, where $n$ is the number of points used to constrain the parameter (for instance for $b$ the number of points in ingress or egress), $\sigma_r$ the level of red noise and $\sigma_w$ the level of white noise ($\sigma_r$ describes the level of systematics that could affect whole portions of the lightcurve). We set $\sigma_r$ to 0.4 mmag in the VLT data and 0.8 mmag in the Magellan data, based on the fluctuations of the residuals. This procedure applied to the combined data yields $b=0 - 0.54$ for the impact parameter, $r/R = 0.110 \pm 0.002$ for the radius ratio and $a/R = 8.07^{+0.44}_{-0.69}$ for the system scale. The uncertainties are higher than in previous determinations using {\em less} accurate data, but we believe they are more realistic and account for the presence of low-level photometric systematics unavoidable in ground-based millimagnitude photometry.

There is an interesting consequence of the fact that the out-of-transit level can cause large biases in the measured transit depth: for a robust determination of the transit depth and therefore of the radius ratio of the system, it is as important to have a robust measurement of the mean flux outside transit as of the transit itself. Given the length of a typical transit, getting both is often impossible from the ground if one wants to sample both ingress and egress. We unwillingly obtained a long stretch of out-of-transit data in our VLT run, due to scheduling constraints, and that may have been crucial in measuring the correct depth.

 \begin{figure}[t]
\centering                     
\includegraphics[width=9.0cm]{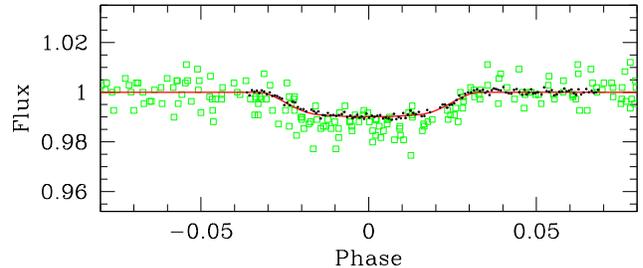}
\caption{For OGLE-TR-56, phased OGLE photometry (open squares), VLT/FORS photometry (dots), and best-fit model transit curve (solid line).}
\label{ogle56}
\end{figure}
\subsection{Planet and star parameters}

By combining the photometric constraints  with radial velocity data (OGLE-TR-10$b$: Bouchy et al. 2005; Konacki et al. 2005; OGLE-TR-56$b$: Bouchy et al. 2005), spectroscopic parameters (Santos et al. 2006, H06)\footnote{$T_{\mathrm eff} =6119 \pm 62$ K, $\log g = 4.21 \pm 0.19$, [Fe/H] =$\ 0.25 \pm 0.08$ for OGLE-TR-56; for OGLE-TR-10 see text.}, and an interpolation of Girardi et al. (2002) stellar evolution models, we derived the new stellar and planetary parameters presented in Table~\ref{param}.  

For OGLE-TR-56, we have not combined the Bouchy et al. (2005) radial velocity measurements with that of Torres et al. (2004), because the unknown zero-point and scale shifts between radial velocity data acquired with different instruments and techniques would introduce more free parameters for little gain in accuracy.

For OGLE-TR-10, we use intermediate values between the parameters of Santos et al. (2006) and H06: $T_{eff}=5960 \pm 100$ and [Fe/H] = 0.15 $\pm$ 0.1. We do not use the $\log g$ constraint. See Section~\ref{sizeogle10} for a discussion of the spectroscopic parameters of OGLE-TR-10.

\begin{table}
\begin{tabular}{l l l } \hline \hline
{\it Star}  & OGLE-TR-10 & OGLE-TR-56\\
 & \\
 Mass [\Msol ] & 1.10 $\pm$  0.05 & 1.17 $\pm$  0.04  \\
 Radius [\Rsol ]& 1.14 $^{+0.11}_{-0.06}$ & 1.32 $\pm$  0.06  \\
 & \\
 {\it Planet} & OGLE-TR-10b & OGLE-TR-56b \\
 & \\
Mass [$M_J$]& 0.61 $\pm$  0.13 & 1.29 $\pm$ 0.12 \\
Radius [$R_J$]& 1.22 $^{+0.12}_{-0.07}$  & 1.30 $\pm$ 0.05  \\
Period [days]& 3.10129  & 1.211909 \\
Transit epoch[BJD] & 2453890.678  & 2453936.598   \\\hline
\end{tabular}
\caption{Parameters obtained in this paper for the OGLE-TR-10 and OGLE-TR-56 systems, host star and transiting planet. The uncertainties on periods and transit epochs are about 1 on the last digit.} 
\label{transit}
\label{param}
\end{table}

\section{Discussion}
\subsection{Systematics in crowded field differential photometry}

Figures~\ref{ogle10} and \ref{ogle56} compare the transit shape derived from our VLT data and H06 with the original data from the OGLE survey.

The original OGLE data for OGLE-TR-10 from Udalski et al. (2002) show a clear discrepancy in transit depth compared to the VLT/FORS lightcurve (lower panel of Figure~\ref{ogle10}). However, the OGLE data for this object exhibit strong night-to-night zero-point variations, to the level of about one percent in flux, due to severe crowding. We applied the detrending algorithm described in Kruszewski \& Semeniuk (2003), using all the objects in the field, to obtain a new systematics-decorrelated OGLE lightcurve (upper panel of Figure~\ref{ogle10}). The agreement between the FORS and OGLE curves becomes very close (the detrended OGLE lightcurve for OGLE-TR-10 is available from the OGLE website). 

There are also systematic effects in the H06 photometry for OGLE-TR-10 that, as explained in Section~\ref{trafit}, affect the out-of-transit mean flux and decrease the apparent transit depth. Gillon et al. (2007) discussed, in the case of OGLE-TR-132, how systematic effects in differential image analysis photometry can cause correlated offsets, without increasing the point-to-point dispersion. A differential image analysis software like ISIS (Alard 1999) tries to find the solution of the following problem: assuming that the stellar flux in the analysed image is the same as in the master image after multiplication by a scaling factor, it must find the background correction and analytical convolution kernel to connect both images. The measured difference flux will be close to the actual flux difference between the analysed and reference images only if the surrounding stars are numerous enough and have a constant photometry relative to each other during the whole run, and if the background can be modelled correctly. Some inaccuracy in the normalisation obtained can be introduced both by intrinsic stellar variability in the field and by differences in atmospheric conditions relative to the reference frame.

In general, the formal uncertainties on transit parameters calculated from transit lightcurves should be taken with caution. As an example, in the case of the transiting system HD189733, using 16 total or partial transit measurements in several filters from different sites and instruments, Bakos et al. (2006a) find that the transit solution from the photometry alone is incompatible with the known radius of the star (the radius of HD189733 is well determined from the Hipparcos parallax, spectroscopy and infrared colours). The authors conclude that systematics in the photometry are responsible for the mismatch. 

Deconvolution photometry requires much more computer time than differential image analysis, but it is more robust towards systematic scale problems, because it does not assume a unique scaling factor between the analysed image and a single reference. It leaves the flux of each star on each image as a free parameter. Moreover, it calculates a full numerical kernel instead of an analytical fit, and allows a finer modelling of the background by detecting faint blended sources down to the noise limit (see Fig.~\ref{fig:a} and the articles cited in the Introduction). It should therefore give very reliable values for the transit depth -- although more subtle systematic effects cannot be excluded entirely.

We note that, despite the difficulties, obtaining reliable transit depth measurements from the ground is possible. In addition to the close agreement between the detrended OGLE and FORS lightcurves for OGLE-TR-10, good agreement for transit depths between OGLE and subsequent measures with large telescopes have also been found for OGLE-TR-111 (Winn et al. 2007), OGLE-TR-113 (Gillon et al. 2006) and OGLE-TR-132 (Gillon et al. 2007).

\subsection{The size of OGLE-TR-10$b$}

\label{sizeogle10}

Our measurements confirm that the transit of \Oten is markedly shallower than indicated by the original OGLE data, while giving a transit depth about 20\% higher than the lightcurve of H06. Thus the planetary radius we derive for OGLE-TR-10b, other parameters being equal,  exceeds by about 10\% the previous estimated by Santos et al. (2006) based on H06, but it does not confirm the very inflated radius calculated from the original OGLE transit lightcurve by Bouchy et al. (2005).

\begin{table}
\centering
\begin{tabular}{ccc}
\hline
 $M_p$ [$M_J$] &  $R_p$ [$R_J$] & Reference\\
\hline 
                    &  $\sim$ 1.1 & Udalski et al. (2002a)\\
0.66 (0.21) & 1.54 (0.12) & Bouchy et al. (2005)\\ 
0.57 (0.12) & 1.24 (0.09) & Konacki et al. (2005)\\  
0.54 (0.14) & 1.06 (0.08) & Holman et al. (2006)\\ 
0.64 (0.14) & 1.14 (0.09) &     Santos et al. (2006)    \\
                    &  1.43 (0.10) & '' \\
                    &                     &  \\
0.63 (0.14) & 1.22 (0.07) & this paper\\ 
\hline
\end{tabular}
\caption{Evolution of the mass and radius derived for OGLE-TR-10$b$.}
\label{tab:d}
\end{table}

There has been a disagreement on the spectroscopic parameters of the host star of \Oten (see Introduction), with Bouchy et al. (2005) and Santos et al. (2006) finding higher temperatures and metallicities ($T_{eff} = 6075 \pm 86$ K, [Fe/H]$\simeq$ +0.3), and Konacki et al. (2005) and H06 lower temperatures ($T_{eff} = 500 \pm 100$ K, [Fe/H] $\simeq$ 0).Both determinations are based on well-established methods, and are compatible at the 3$\sigma$ level. For other OGLE targets, these two groups have found compatible results. We have not attempted to resolve these differences here, and we adopted intermediate values of $T_{eff} = 5960 \pm 100$ K and [Fe/H]$ = 0.15 \pm 0.1$ for our calculations. The temperature, combined with stellar evolution models, is the main constraint on the mass of the primary. Since the primary mass enters the transit equations to the $1/3^{th}$ power only, the uncertainties in the temperature do not dominate the final error budget.

\begin{figure}

\centering                     
\includegraphics[width=9.0cm]{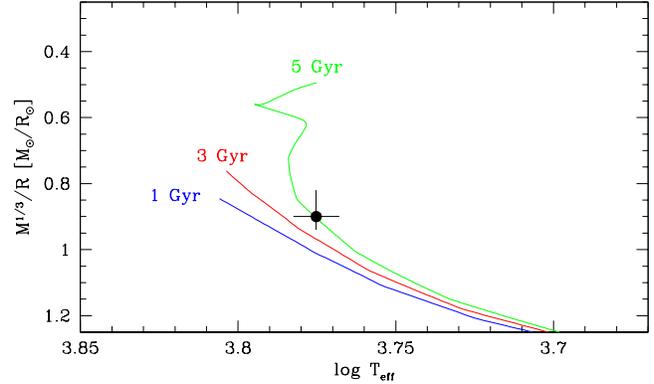}
\caption{Position of OGLE-TR-10 and the Pietrinferni et al. (2004) stellar evolution tracks for [M/H]=0.06 in the temperature vs. $M^{1/3}/R$ plane, at ages 1, 3 and 5 Gyr. The position of OGLE-TR-10 is plotted with our adopted spectroscopic temperature and $M^{1/3}\! /R$ from the transit shape. }
\label{fig:jf}
\end{figure}

H06 show that the spectroscopic parameters of Santos et al. (2006) place OGLE-TR-10 below the ZAMS in the temperature-gravity diagram. Gravity is a difficult spectroscopic parameter to constrain precisely from spectra, and in the case of transit parameters, the main constraint on the primary radius does not come from the gravity measurements but from the constraint on the $M^{1/3}/R$ ratio set by the transit duration and shape. Therefore it is best to visualize the comparison of the data with stellar evolution models in the plane of the two strongest constraints on the properties of the primary, temperature and $M^{1/3}/R$. Fig.~\ref{fig:jf} gives the position of OGLE-TR-10 in this plane, showing that both temperature determinations are compatible with stellar evolution models for reasonable ages of a few Gyr.

Table~\ref{tab:d} summarises the evolution of the radius determination for \Oten. Perhaps ironically, our value for the radius of  \Oten turns out to be near the average of all previously published values weighted by their errors.%, showing that - sometimes - statistics works.

\subsection{The size of OGLE-TR-56$b$}

Our lightcurve for OGLE-TR-56 shows a shallower transit than the original OGLE data (radius ratio 0.101 $\pm$ 0.002 instead of 0.114 $\pm$ 0.004). In this case, the OGLE lightcurve was already detrended with the Kruszewski \& Semeniuk (2003) method, but there may be some remaining systematics due to crowding (see Figure~\ref{fig:a} and previous Section). The transit ingress and egress are relatively long and well defined in the FORS data, so that the impact parameter can be measured with precision. The shallower transit implies a smaller radius ratio, but the higher impact parameter requires a larger primary, so that the planetary radius turns out only slightly larger than found by Santos et al. (2006) with the OGLE lightcurve and the same spectroscopic parameters.Because the transit occurs at a very high impact parameter, low uncertainties in $b$ translate into high uncertainties on the primary and planet radius. In fact, if we include a moderate amount of red noise described by $\sigma_r=0.4$ mmag, then the uncertainty on the planetary radius becomes 0.19 R$_J$, so that radius values as high as 1.5 R$_J$ are not entirely excluded. This illustrates the difficulty of lifting the degeneracy between orbital angle and primary radius in transit lightcurves with ground-based data. Even with lightcurves of very high quality, systematics well below the millimagnitude level are sufficient to introduce significant errors in the measurement of the planetary radius.

\subsection{Impact on Hot Jupiter structure and evolution}

The VLT lightcurve data confirm both \Ofive and \Oten as inflated hot Jupiters, together with HD 209458$b$ and the newly discovered HAT-P-1$b$ (Bakos et al. 2006b) and WASP-1$b$ (Collier Cameron et al. 2006).   

\Ofive being much more massive than other inflated hot Jupiters, is a  challenging case to model. Not only is it the planet with the shortest period discovered to date, but it also has a high "missing" energy problem (Guillot et al. 2006), since its outer parts must be inflated against a stronger gravity. Its large size also makes it more vulnerable to evaporation (Lecavelier et al. 2006).  

A large radius could be explained if \Ofive is very young and has not contracted yet to its asymptotic radius value  (gas giant planets are expected to contract with time). However, Melo et al. (2006) have examined the issue of the age of OGLE-TR-56, and find that it has to be older than 0.5~Gyr, from several different age indicators. In most models, 0.5~Gyr is sufficient for hot Jupiters to contract to a radius near their asymptotic value.

Determinations of the radius of OGLE-TR-10 have varied widely since its discovery (see Table~\ref{tab:d}), spanning all the range from the largest gas giant to the densest. We believe we have now converged to a robust value with a satisfactorily low uncertainty. The position of \Oten in the mass-radius diagram is close to that of HAT-P-1$b$ and HD 209458$b$.  The case of OGLE-TR-10$b$ is critical in the relation proposed by Guillot et al. (2006) between the stellar metallicity and planetary radius. Our new value of the radius of OGLE-TR-10$b$ makes it closer to the rest of the transiting planets in the parameter space ``radius anomaly'' versus star metal content. 

%This study shows how refining the parameter determination for known transiting hot Jupiters is as important as discovering new transiting planets for the understanding of gas giant structure and evolution.

Following the recent precise high-accuracy photometry for OGLE-TR-111$b$ (Winn et al. 2006), OGLE-TR-113$b$ (Gillon et al. 2006) and OGLE-TR-132$b$ (Gillon et al. 2007), the present study on OGLE-TR-10$b$ and OGLE-TR-56$b$ completes the precise radius determinations for the five known transiting planets from the OGLE survey.

\begin{acknowledgements} 
The authors thank the ESO staff at Paranal for their diligent and competent execution of the observations, as well as Matthew Holman, Joshua Winn and Joel Hartman for stimulating discussions. This paper was much improved by a very detailed reviewing from an anonymous referee. Support from the Funda\c{c}\~ao para a Ci\^encia e a Tecnologia (Portugal) to N.C.S. in the form of a fellowship (reference SFRH/BPD/8116/2002) and a grant (reference POCI/CTE-AST/56453/2004) is gratefully acknowledged. DM, WG, GP and MTR gratefully acknowledge financial support for this work from the Chilean Center of Astrophysics FONDAP 15010003. G.P. and A.U. were partly supported by the Polish MNSW DST grant to the Warsaw University Observatory.
\end{acknowledgements}

\end{document}